# State of the Art of Augmented Reality (AR) Capabilities for Civil Infrastructure Applications

Jiaqi Xu[1], Derek Doyle[2], Fernando Moreu[3*]

[1] *Postdoctoral Associate, Department of Civil, Construction & Environmental Engineering (CCEE), University of New Mexico (UNM), Albuquerque, NM, USA. Email: xujiaqi@unm.edu*

[2] *Assistant Chief Scientist, Air Force Research Laboratory, Space Vehicles Directorate, Kirtland Air Force Base, Albuquerque, NM, USA. Email: derek.doyle@us.af.mil*

[3] *Assistant Professor, CCEE; Assistant Professor (courtesy appointment) Electrical and Computer Engineering; Assistant Professor (courtesy appointment) Mechanical Engineering; Assistant Professor (courtesy appointment) Computer Science, UNM, Albuquerque, NM, USA. Email: fmoreu@unm.edu*

[*] *Corresponding author.*

## Abstract

Augmented Reality (AR) is a technology superimposing interactional virtual objects onto a real environment. Since the beginning of the millennium, AR technologies have shown rapid growth, with significant research publications in engineering and science. However, the civil infrastructure community has minimally implemented AR technologies to date. One of the challenges that civil engineers face when understanding and using AR is the lack of a classification of AR in the context of capabilities for civil infrastructure applications. Practitioners in civil infrastructure, like most engineering fields, prioritize understanding the level of maturity of a new technology before considering its adoption and field implementation. This paper compares the capabilities of sixteen AR Head-Mounted Devices (HMDs) available in the market since 2017, ranking them in terms of performance for civil infrastructure implementations. Finally, the authors recommend a development framework for practical AR interfaces with civil infrastructure and operations.

**Keywords**: Augmented Reality (AR); Head-Mounted Devices (HMDs); civil infrastructure; human infrastructure interfaces; capabilities; classification.

## 1. Introduction

Augmented Reality (AR)/Virtual Reality (VR) devices are attracting both technology giants and startups dedicated to both hardware and user-friendly software tools. According to the analysis of the International Data Corporation conducted in November 2019, the global AR/VR market will reach $18.8 billion USD in 2020 [1]. AR technologies are now moving from research prototypes to applications in practical projects [2], evolving from futuristic visions to engineering tools. Today, AR aims to transform medicine, defense, education, and the interaction between humans and engineering in design, manufacturing, and product management [3].

AR displays the real environment enhanced by interactive virtual (computer graphic) objects. AR has enabled new opportunities in emergence of large and complex data, contributing with new theories and applications [3–5]. The concept of AR has been outlined by Caudell and Mizell in 1990 [6]. The concept of AR is often confused with three other developments in computer graphics visualization: VR, Augmented Virtuality (AV) and Mixed Reality (MR). VR is an artificial environment which is experienced through sensory stimuli provided by a computer. VR does not necessarily include interaction with the real environment. AV merges real-world objects onto a virtual environment. The main difference between AR and AV is the focus of the major displaying environment. AR enables more opportunities of quantification of the external world, higher user interaction and exploration with their nearby environment, and wider see-through characteristics [7–15]. In this paper, the authors use AR as an inclusive term that enables engineers interact with infrastructure in real environments through virtual interfaces.

AR implementations in civil systems promote comprehension of the working context during experiments, allow engineers to perform field tasks with awareness of both the physical and synthetic environments, reduce heavy workload of conducting damage detection manually, and provide insight of the dangers on site [16]. About 50% of today's AR implementations in civil infrastructure are focused on construction. In 2010, Dong and Kamat [17,18] developed the SMART-ARMOR system. SMART is an AR authoring language with a standard AR development environment. ARMOR is a mobile computing hardware framework. The SMART-ARMOR system has been tested with a multistory steel frame example, which is a milestone holistic and applicable outdoor construction application of AR technologies. Golparvar-Fard et al. [19–22] applied the registration method to solve the discrepancy check problem. Since 2009 they have provided a platform named $D^4AR$ that shows discrepancy information during construction. In 2013 they updated their $D^4AR$ to the $HD^4AR$ system [23], enabling applications of camera-equipped mobile devices already available. In 2014, based on the $D^4AR$ system, Karsch et al. [24] combined a set of AR tools to build a valuable platform, ConstructAide, to monitor construction progress with unordered photo collections and 3D building models.

High-efficiency collaborative visualization and communication among multiple users can be realized by applying AR technologies. Tele-communication can provide high-performance guidance for workers. Dong et al. [25] developed an AR software, ARVita, for the collaborative visualization of multiple users wearing Head-Mounted Devices (HMDs). Using ARVita, all the users can observe and interact with the dynamic visual simulations of the engineering processes. Hammad et al. [26] developed a prototype distributed AR visualization collaborative system, DARCC, and tested it in a bridge deck rehabilitation project to show the effectiveness. Another significant application is to guarantee site safety by visualizing complex workplace situations with AR devices. For example, Kim et al. [27] developed a vision-based hazard avoidance system to provide safety information for workers. Fang et al. [28] deployed AR technologies to prevent workers operating at hazardous heights. The authors are also planning to contribute to the topic of implementing AR HMDs to improve site safety for engineers and workers in a future research project.

In summary, research groups have developed and designed AR prototype devices and applications in civil infrastructure. However, these successful efforts are in general designed for research specific purposes, and are difficult for civil engineers to implement in their own projects. Different approaches need to be developed in order to increase the growth of AR hardware and software suites across the civil engineerin field. For example, the lack of hardware continuity can be a hurdle to AR implementation that should be considered in assessing utility of tools [29,30]. In this context, Commercial Off-The-Shelf (COTS) solutions that are robust enough to perform specialized tasks across multiple domains and environmental situations. Additionally, COTS solutions should have a diverse enough portfolio to enable adoptions by various industry/projects. This viability is critical to any risk adverse industry's assessment of maturity and reliability via support longevity versus the cost of integrating the technology into the business model and sustaining the learning curve.

In the past, researchers have classified hardware and software capacities of AR devices to increase their understanding and implementation by the engineering community. Soto [16] classified and compared AR devices produced before 2017, and summarized their potential for implementation. The detailed comparison of AR devices provided information for civil engineers to apply them in construction. Soto's results showed that AR HMDs can improve users' visualization for construction steps more effectively than hand held devices and other interfaces. Finally, Soto's study recommended conducting further studies to continue increasing the implementation of AR HMDs in civil infrastructure projects. Today, the comparison from Soto in 2017 needs to be expanded and increased since the majority of today's AR devices have been produced during the last three years and are not included in Soto's study. Additionally, AR devices today have more powerful sensors, more display pixels and faster processing units than previous versions. For example, DAQRI, ODG R-9 and Glassup F4 visor were considered advanced in 2017 by Soto, but today these devices are no longer available.

This paper classifies AR technology capabilities in the context of field implementation for civil infrastructure projects. First, the authors outline the milestones of AR technology from 1968 to date and compare them with the number of publications on that period of time. Then, the authors classify the categories of AR devices that are of higher interest to civil infrastructure applications. This classification enables civil infrastructure professionals to understand the state of the art of today's AR hardware and software. This study compares the technical advantages, disadvantages and limitations of sixteen AR HMDs. The authors consider civil infrastructure areas of interest and rank the sixteen AR HMDs in terms of their general properties, sensors, computational capabilities, and display capabilities. The result of this study is a new classification of AR devices available since 2017 in the context of field implementation.

## 2. Chronology of AR milestones

In the last fifty years, AR research and technology adoption have grown in parallel with the advancements of computer science and interfaces between humans, computers and physical systems. Figure 1 shows the number of publications since the first AR application in 1968 to date, and the most significant AR milestones in the same period of time.

The selection of significant AR milestones is outlined chronologically to better show the historical context of the current state of AR technologies. In 1968, Ivan Sutherland developed a VR HMD with the name 'Sword of Damocles' to show images comprising virtual environment of simple wireframe rooms with head-tracking feature [31]. In 1974, Myron Krueger [32] created a milestone 'Videoplace' laboratory based on his previous developments of 'Glowflow', 'Metaplay' and 'Psychic space'. In the Videoplace laboratory, AR displays were designed to surround the users and respond to the movements and actions of the users. These remarkable AR platforms were developed before the term 'Augmented Reality' was formally coined. In 1990, the term 'Augmented Reality' was introduced formally for the first time by Caudell and Mizell [6]. In 1992, Louis Rosenberg [33] developed 'Virtual Fixture' in the USAF Armstrong Labs, which was the first functional AR platform enabling people to control robot arms by moving physical arms. In 2000, Kato [34] created an open-source software library 'ARToolKit', which has been applied in web browsers in 2009, enabling researchers to share and develop AR tools in the same platform. On April 15, 2013, Google started to sell a Google Glass prototype in the United States, before it became available to the general public on May 15, 2014 [35]. In 2017, Google stopped producing the Google Glass prototype for the public and turned to the enterprise edition [36]. Also, Microsoft announced their AR prototype in March 2016 named Microsoft HoloLens. Microsoft announced the global expansion and commercialization of Microsoft HoloLens in October 2016 [37]. In addition, in 2019, Facebook acquired and reorganized Oculus, becoming Facebook Technologies [38] and is currently considering purchasing ODG [39]. Since then, Facebook has increased their AR technical team [38]. Apple is expected to announce and launch Apple Glasses in 2023 [40]. Currently, Mojo Vision is developing the next generation AR lens by inputting an AR screen in the medical-grade contact lens instead of glasses [41]. However, the publication time for the AR lens has not been announced yet.

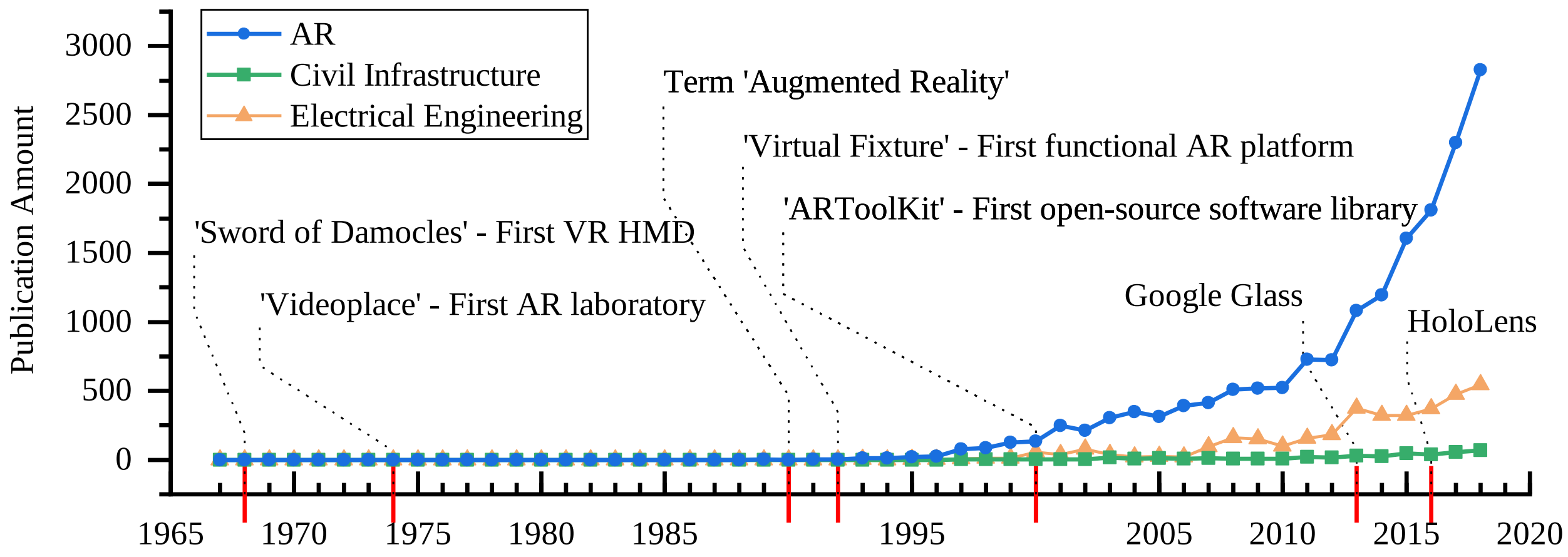

**Figure 1. Number of AR implementation papers included in *Web of Science.***

Figure 1 shows the growth of AR publications since the late 1960s to date. The number of AR publications started to grow in the late 90s. Since then AR publications increased slowly, transforming in 2010 to an exponential growth. In 2018, there were 2825 research papers published regarding the topic of AR on the *Web of Science*. Publications in the field of electrical engineering have followed this trend (red line in Figure 1), whereas publications in civil infrastructure (green line in Figure 1) have remained low. In 2018, 67 articles were published discussing AR applications in civil infrastructure (only 2.4% of the total number of AR implementations). In the same year, 546 publications (19.3% of the total) were focused on AR applications in electrical engineering, almost eight times higher. Civil infrastructure has fallen behind in both theory and application of AR technologies. It would be valuable to increase interactions between humans and infrastructure [4]. Practitioners in civil infrastructure are averse to innovations unless the capabilities and success to date is documented and shared by the community. The next section of this paper summarizes basic AR advances for potential adopters in the civil infrastructure field who may not familiar with computer science in their regular practice.

## 3. Categories of AR devices

This paper classifies the AR devices that are commercially available since 2017 until February 2020 for civil infrastructure implementations. Fully immersive VR headgear or video viewers are not included, because the emphasis of this study is the ability of engineers to interact with real environments in the field. There are three categories of AR devices based on different display methods, i.e., HMDs, handheld display, and spatial display (Figure 2). In this section, the authors analyze the specific contributions and capabilities of the three types of AR display methods in terms of their advantages and limitations in the context of civil infrastructure implementation.

HMDs consist of smart glasses displaying virtual images superimposed in front of the user's eyes. Based on the number of displaying projectors, HMDs can be categorized into either Monocular-see-through or Binocular-see-through. Binocular-see-through is more popular than Monocular-see-through by better fitting people's observational intuition. Some HMDs are designed as a notification system, being controlled by a tablet or controller. For example, Toshiba dynaEdge AR100 Smart Glasses is controlled by Toshiba dynaEdge DE-100-12U Mobile Mini PC, and Glassup F4 visor is controlled by an external controller. Although these devices have a controller PC/tablet, they should still be classified as HMDs. HMDs give solutions to initial technical difficulties such as crosstalk of the backside imager [42], scene recognition and natural features tracking [43–45], and close-up view coherent augmentation [46]. Limitations on HMDs include concerns about safety as the user can be isolated from tripping hazards or other obstacles in the field [47,48].

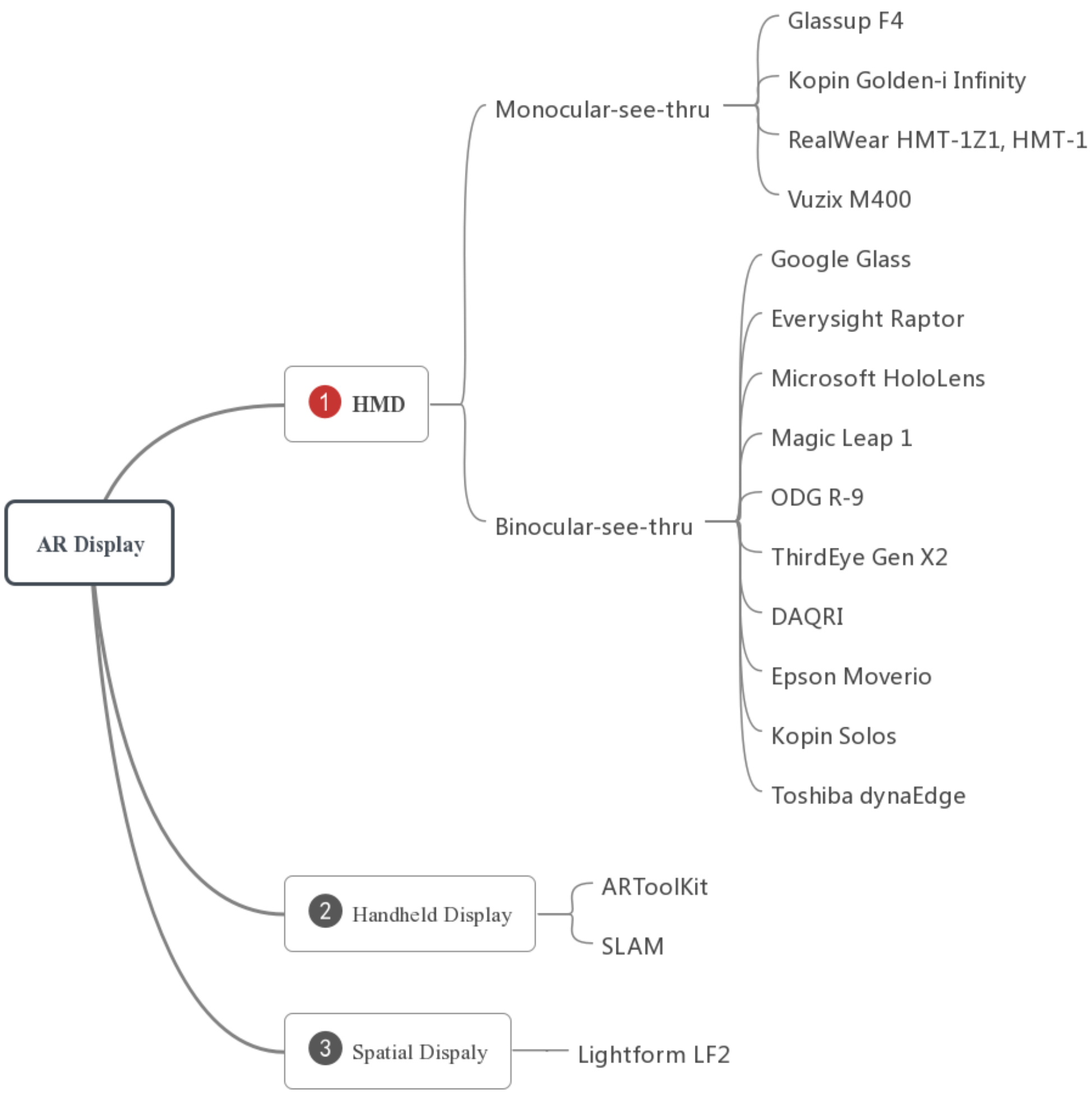

**Figure 2. Categories of AR devices in terms of display.**

Handheld displays are related to AR applications developed for handheld devices. These displays show virtual images on handheld computing devices, e.g., smart phones, tablets, Personal Digital Assistants (PDAs), etc. The handheld display is typically represented by an application with an AR function, e.g., Morpholio AR Sketchwalk [49], GAMMA AR [50], ViewAR [51], etc. The comparison of updated AR applications can be found in the SocialCompare website [52]. The advantages of the handheld displays are the increased awareness of the real environment and confinement of virtual elements in platform that isn't intended to be in your field of view (FoV) at all times, while the limitations are that users' hands are not available for other activities in the field, compromising their safety [47,48].

As an alternative to HMDs and handheld devices, spatial display devices show virtual images directly in the environment. The advantages of spatial display are that the safety of users is not undermined by the AR display and multiple observers can utilize a single set of hardware, as opposed to everyone using one dedicated device per person, either in their hands or on their heads. Lightform Company, founded in 2014, develops high-resolution AR projections that track objects and respond to human inputs in real-time [53]. Their first and second generation products, LF1 and LF2 were presented in February and November of 2019, respectively. However, the lack of portability of spatial devices limits their applications in the field of assessment and inspection of large infrastructure sites. In addition, the complexity of augmented elements is typically more confined to 2D overlays that require existing structure present.

In conclusion, HMDs are the most commonly implemented display method for professional applications. Most AR devices in the market today are HMDs, but the unclassified complex information has limited the mass adoption by civil engineers. In the following section, the authors analyze and compare sixteen AR HMDs in the last three years in the context of potential for civil infrastructure field implementation.

## 4. Comparison of AR HMDs

The foundation of AR implementation is the theoretical development of AR technologies. Most devices today are in the experiment and prototype phase, and have not been implemented in real site constructions. Before 2017, the main difficulties of adopting AR technologies in civil infrastructure have been the high expense and the immaturity of devices [54]. With the rapid development of AR devices in recent years, commercially available AR HMDs with reasonable weight and price have been developed for civil infrastructure. Although a perfect final theoretical solution to the AR implementation is still in the distant future, AR HMDs are using alternatives to meet the engineering demands. As compared in Tables 1-4, a suitable AR HMD should be selected by first considering performance requirements for different engineering scenarios, then choosing along the price/weight line in Figure 5. For example, the RealWear HMT-1Z1 should be selected for high-risk (e.g., potentially explosive sites) and high-value construction for its product durability and high price; the Google Glass Enterprise Edition 2 is more suitable for daily use because of its light weight and ergonomic design.

### 4.1. General properties

Table 1 summarizes the general capabilities of AR HMDs, sorted by released date from #1 (most recent, February 2020) to #16 (January 2017). AR HMDs #1 and #16 are Vuzix M400 Version 1.1.4 and DAQRI Smart Glasses, respectively. AR HMDs #14-16 are no longer available for purchase: Glassup F4 visor, ODG R-9 and DAQRI, respectively. Out of the total sixteen AR HMDs, eleven are developed by United States companies (69%), and three are produced by two Japanese companies (19%): Toshiba and Epson.

According to Golparvar-Fard et al. [55], product durability is among the first considerations when selecting AR devices for engineering applications. Table 1 shows four indexes for product durability ($5^{th}$-$8^{th}$ columns): water resistance, working temperature range, working humidity range, and drop safe distance. According to these four columns of Table 1, the Vuzix M400 Version 1.1.4 and RealWear HMT-1/1Z1 are the two AR HMDs with highest product durability.

The $9^{th}$ and $10^{th}$ columns of Table 1 list weight and price for each AR HMD, respectively. Figure 3 shows the weight distribution of AR HMDs from 2017 until today, with a median weight value of 218g per device. The median value is used to avoid the influence of devices with excessive weight such as HMT-1Z1, DAQRI and Microsoft HoloLens 2. According to Figure 3, the distribution of weight of AR HMDs does not follow a chronological trend. Weight limits should be considered when selecting AR HMDs in the field. The American National Standards Institute (ANSI) Z89.1-1969 regulates the maximum weight limit for safety hard hats used in civil infrastructure industry [56]. According to ANSI, Class A and C hard hats have a maximum recommended weight of 425g, and Class D hard hats has a maximum recommended weight of 850g. AR HMDs should follow the ANSI regulation regarding maximum weight limit in the civil infrastructure industry. The majority of the AR HMDs shown in Figure 3 weight less than the limit of Class A/C hard hats. The weight of HMT-1Z1 and Microsoft HoloLens 2 are beyond the limit of Class A/C hard hats, but still under the limit of Class D hard hats. It can be concluded that the majority of AR HMDs are in general available for implementation in the field of civil infrastructure, although according to experts other specific field requirements besides weight such as safety and comfort still need to be considered [56].

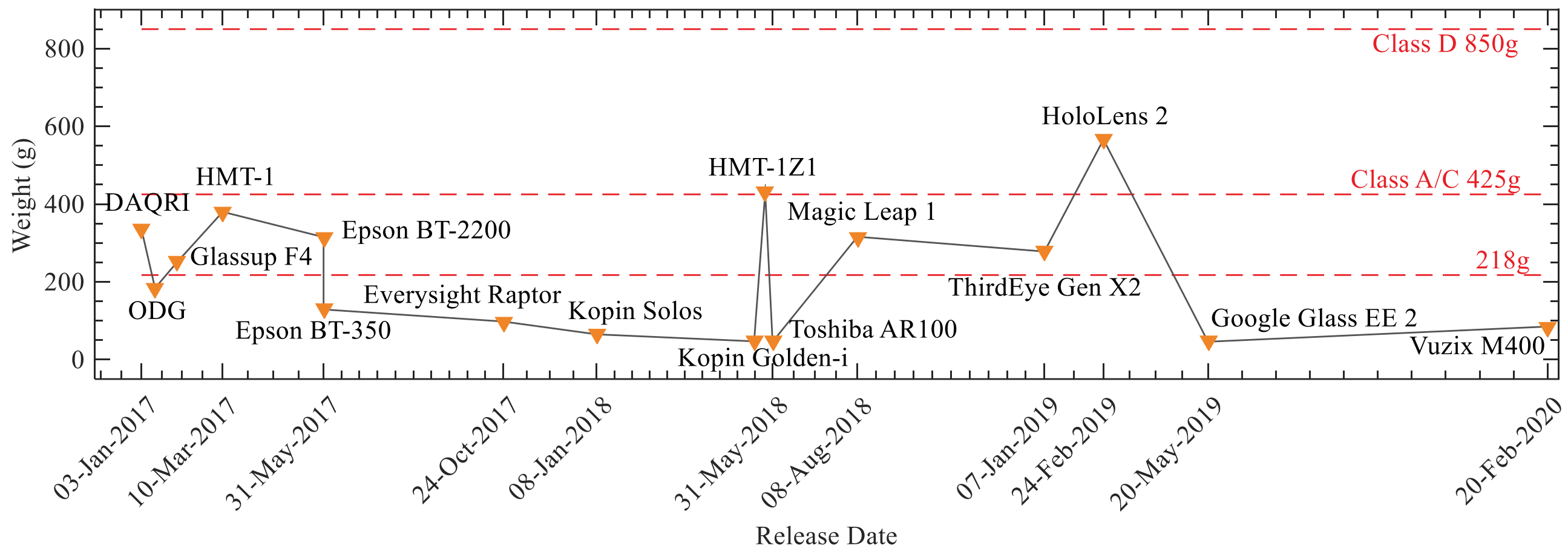

**Figure 3. Chronological trend of AR HMDs weight (01/2017-02/2020).**

Figure 4 shows the price distribution of the AR HMDs. According to this analysis, the median price value per device is $1975. As in the case of weight, the change of AR HMDs cost between 2017 and 2020 does not follow an evident chronological trend.

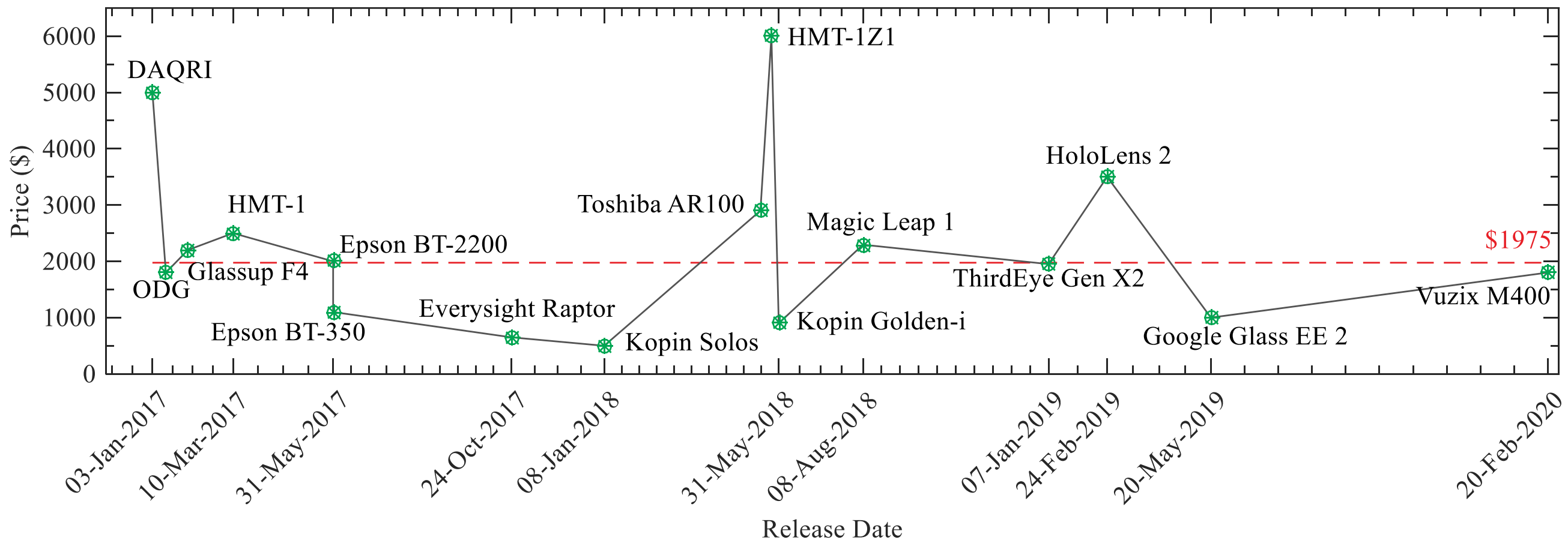

**Figure 4. Chronological trend of AR HMDs price (01/2017-02/2020).**

Figure 5 shows the price distribution of the AR HMDs versus weight. With the exception of three devices (Toshiba AR100, DAQRI and HMT-1Z1), the other thirteen AR HMDs follow a linear correlation between price and weight with a regression index *r*=0.91. This linear correlation is of interest to potential adopters of AR HMDs to understand available product capabilities and limitations in weight and cost. HMT-1Z1 and DAQRI are the two most expensive AR HMDs: HMT-1Z1, priced at $6000, is developed by RealWear, a startup company founded in 2015, and it is the only AR HMD especially developed for an outdoor environment. DAQRI is one of the earliest AR HMD trials and is no longer available for purchase. According to past research, heavier AR HMDs are generally attributed to larger processing units and sophisticated add-on hardware, which enable more advanced user experiences and better performances [57]. In addition, this mass can play into durability and thermal management of the hardware to survive more extreme environments. After AR HMDs are established as products of widespread interest to engineers, they are expected to decrease price and weight while increasing their performance. Civil engineers and other infrastructure industries can actively participate in the direction of development of AR HMDs and technologies.

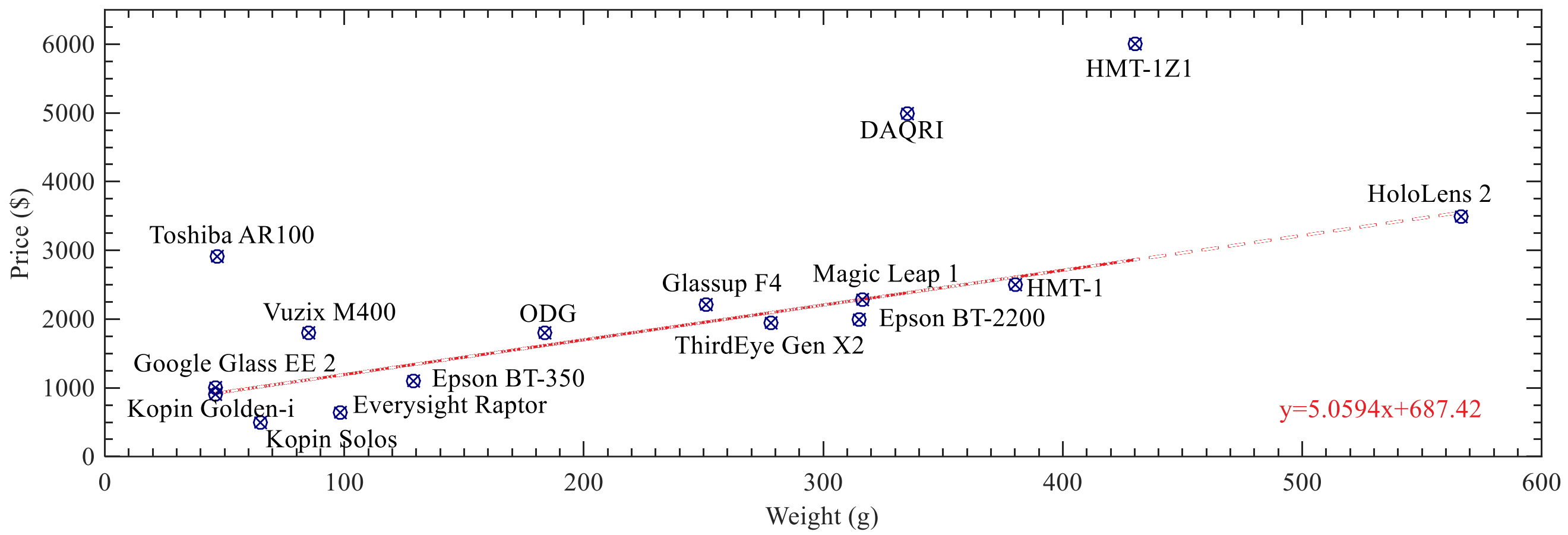

**Figure 5. AR HMDs linear correlation between weight and price (01/2017-02/2020).**

### 4.2. Sensors

Table 2 summarizes the sensors built in AR HMDs, sorted by the number of pixels on its camera sensor. The 3rd column in Table 2 summarizes the pixels of cameras of the sixteen devices. AR HMDs with more pixels can capture more accurate images. HMT-1/1Z1 from RealWear has the most advanced cameras of 16Mp, with good accuracy for infrastructure inspections and data acquisition.

Tracking characteristics include head tracking and eye tracking, as shown in the 4th and 5th column of Table 2, respectively. Head-tracking is a necessity for AR HMDs to properly align the rendering of virtual objects onto real environments [58]. Therefore, all the AR HMDs available for purchase today include the function the head-tracking. Eye tracking enables AR HMDs to capture the eye position and improve the occlusion accuracy of virtual objects [59]. The ThirdEye Gen X2, Vuzix M400 Version 1.1.4, Google Glass Enterprise Edition 2, Microsoft HoloLens 2, and Magic Leap 1 include the eye tracking function. HMDs' ability to precisely depict the real environment depends on the mapping capacity enabled by depth sensors. In the devices with eye tracking function, only ThirdEye Gen X2, Microsoft HoloLens 2, and Magic Leap 1 are developed with depth sensors. Gyroscope, accelerometer, magnetometer and locating sensors enable the HMDs to be implemented in outdoor environments. Microsoft HoloLens 2 is a functional device for indoor environments due to the lack of locating sensors.

### 4.3. Computational capacities

One of the major challenges of AR implementation is the synchronization problem, i.e., time delay problem. The images in the real environment are captured by the users' eyes immediately, but the images in the virtual environment have to be processed before displayed in front of eyes. The time delay caused by the processing of the virtual objects can lead to a displacement between the real and virtual environment [60]. For the optical see-through HMDs, the synchronization cannot be thoroughly eliminated, but can be minimized by accelerating the virtual objects processing and improving computational capacities.

Table 3 summarizes the computational and connectivity capacities of AR HMDs, sorted by the Random Access Memory (RAM) and hardware storage, listed in the 4th and 5th columns, respectively. For professional work, 16GB RAM is ideal, and 4GB is an entry-level memory to produce applicable software. The processing unit determines the computational capacity of a device, as listed in the 3rd column of Table 3. The two devices developed by Kopin, Golden-i Infinity Smart Screen and Solos Smart Glasses, and the AR projector, Lightform LF2, use the processor and internal storage of the host device (computer or smart phone). Except for the Kopin and Lightform products, over half of the AR HMDs are developed with a Qualcomm Snapdragon processor. Intel cores are the second-most popular processors.

The 6th-8th column of Table 3 list the connectivity capacities of the AR HMDs. Universal data transfer modes include WiFi, Bluetooth and USB. Nearly all the devices are designed to have all the three universal data transfer ports. Battery life is another important index for AR HMDs, measured by the interval between two full charges. Nevertheless, besides the battery life, continuous work time also depends on the use intensity and therefore can't be simply measured. Comparatively, Microsoft HoloLens 2 has the maximum battery capacity, but Everysight Raptor and Kopin Solos Smart Glasses have longer active use time because of their relatively power saving design.

### 4.4. Displays

Table 4 summarizes the display capacities of the AR HMDs, sorted by the FoV which is listed in the 3rd column. FoV is an index describing the open see-through angle of an optical device. A larger FoV provides engineers a broader observable environment. In Table 4, FoV is represented by the diagonal see-through angle. The 5th column of Table 4 lists the resolution of AR HMDs, defined as the number of distinct pixels that can be displayed in each dimension. The 6th column of Table 4 compares the refreshing rate, defined as the number of display images a device can update per second. A trade-off exists between the resolution and the refresh rate [61]. A higher resolution enhances display quality, while a higher refresh rate improves video fluency [62]. Both Google Glass Enterprise Edition 2 and Kopin Solos compromise resolution to ensure a refresh rate of 120 Hz, while the Everysight Raptor and the Magic Leap 1 have a decreased refresh rate to fulfil the 1080p resolution.

### 4.5. Summary

This section provides a systematic classification of available AR HMDs regarding to parameters of interest to civil engineers. Table 1 compares the weight, price and product durability of sixteen AR HMDs. The lightest three AR HMDs are Kopin Golden-i, Toshiba AR100 and Kopin Solos. The least expensive three AR HMDs are Kopin Solos, Everysight Raptor and Kopin Golden-i. HMT-1Z1 has the greatest product durability. Table 2 compares the sensors of the AR HMDs. According to this table, HMT-1/1Z1 have the largest camera resolution. Table 3 compares the computational capabilities of the AR HMDs. Magic Leap 1, ODG R-9 and Vuzix M400 Version 1.1.4 have the largest RAM and storage. Table 4 compares the display capabilities of the AR HMDs. Google Glass Enterprise Edition 2, Everysight Raptor and Microsoft HoloLens 2 have the largest FoV. Everysight Raptor, Magic Leap 1 and ODF R-9 have the largest resolution. Google Glass Enterprise Edition 2, Microsoft HoloLens 2 and Kopin Solos have the largest refresh rate. The classification and ranking of the capabilities in these four tables are focused on implementations for civil infrastructure applications, but are not exhaustive and other areas can be added in the future. This classification is however a first step to enable the civil infrastructure community to choose AR devices highly ranked in the areas of interest for their specific implementations. This classification is also a comparative list of options where they can better understand the tradeoff of their selection. The four tables enable civil infrastructure potential adopters a simple to read list of technical capabilities related to civil infrastructure applications.

**Table 1. Chronology and general properties of recent AR HMDs.**

| # | Device | Release Date | Country | Water Proof | Working Temperature | Relative Humidity | Drop Safe | Weight | Price |
|---|---|---|---|---|---|---|---|---|---|
| 1 | Vuzix M400 Version 1.1.4 [63] | Feb-20-2020 | U.S. | IP67 | 0-45℃ | 0-95% | 2m | 85g | $1,800 |
| 2 | Google Glass Enterprise Edition 2 [64] | May-20-2019 | U.S. | IPX3 | 5-45℃ | - | - | 46g | $999 |
| 3 | Microsoft HoloLens 2 [65] | Feb-24-2019 | U.S. | - | 10-27℃ | 8-90% | - | 566g | $3,500 |
| 4 | ThirdEye Gen X2 Mixed Reality Smart Glasses [66] | Jan-07-2019 | U.S. | - | - | - | 2m | 278g | $1,950 |
| 5 | Magic Leap 1 [67] | Aug-08-2018 | U.S. | - | 10-25℃ | - | - | 316g | $2,295 |
| 6 | Kopin Golden-i Infinity Smart Screen [68] | May-31-2018 | U.S. | IP67 | - | - | 2m | 46g | $899 |
| 7 | RealWear HMT-1Z1 [69] | May-31-2018 | U.S. | IP66 | -20-50℃<br>Intrinsically safe | - | 2m | 430g | $6000 |
| 8 | Toshiba dynaEdge AR100 Head Mounted Display [70] | May-16-2018 | Japan | IP53 | -20-60℃ | 0-95% | - | 47g | $2,900 |
| 9 | Kopin Solos Smart Glasses [71] | Jan-08-2018 | U.S. | - | - | - | - | 65g | $499 |
| 10 | Everysight Raptor [72] | Oct-24-2017 | Israel | IP55 | 0-40℃ | 5-95% | - | 98g | $649 |
| 11 | Epson Moverio BT-350 Smart Glasses [72] | May-31-2017 | Japan | - | 5-35℃ | 20-80% | - | 129g | $1,099 |
| 12 | Epson Moverio Pro BT-2200 Smart Headset [73] | May-31-2017 | Japan | IP54 | 0-40℃ | 20-80% | 1.2m | 315g | $1,999 |
| 13 | RealWear HMT-1 [74] | Mar-10-2017 | U.S. | IP66 | -20-50℃ | - | 2m | 380g | $2500 |
| 14 | Glassup F4 visor [75] | Feb-2017 | Italy | IP31 | 5-35℃ | - | - | 251g | €2000<br>($2200) |
| 15 | ODG R-9 [76] | Jan-2017 | U.S. | - | - | - | - | 184g | $1,800 |
| 16 | DAQRI Smart Glasses [77] | Jan-2017 | U.S. | - | - | - | - | 335g | $4,995 |

**Table 2. Sensors of recent AR HMDs.**

| # | Device | Camera | Head Tracking | Eye tracking | Depth | GAM | Location |
|---|---|---|---|---|---|---|---|
| 7 | RealWear HMT-1Z1 [69] | 16Mp | √ | - | - | √ | GPS/GLONASS/A-GPS |
| 13 | RealWear HMT-1 [74] | 16Mp | √ | - | - | √ | GPS/GLONASS/A-GPS |
| 10 | Everysight Raptor [72] | 13Mp | √ | - | - | √ | GPS/GLONASS |
| 6 | Kopin Golden-i Infinity Smart Screen [68] | 13Mp | √ | - | - | √ | GPS/GLONASS/A-GPS/GALILEO |
| 4 | ThirdEye Gen X2 Mixed Reality Smart Glasses [66] | 13Mp | √ | √ | √ | √ | GPS |
| 1 | Vuzix M400 Version 1.1.4 [63] | 8Mp | √ | √ | - | √ | GPS/GLONASS |
| 2 | Google Glass Enterprise Edition 2 [64] | 8Mp | √ | √ | - | √ | GPS/GLONASS |
| 3 | Microsoft HoloLens 2 [65] | 8Mp | √ | √ | √ | √ | - |
| 15 | ODG R-9 [76] | 8Mp | - | - | - | √ | GNSS w/ iZAT |
| 16 | DAQRI Smart Glasses [77] | 8Mp | √ | - | √ | √ | GPS |
| 8 | Toshiba dynaEdge AR100 Head Mounted Display [70] | 5Mp | √ | - | √ | √ | GPS |
| 5 | Magic Leap 1 [67] | - | √ | √ | √ | √ | GPS |
| 11 | Epson Moverio BT-350 Smart Glasses [72] | 5Mp | √ | - | - | √ | GPS |
| 12 | Epson Moverio Pro BT-2200 Smart Headset [73] | 5Mp | √ | - | √ | √ | GPS |
| 14 | Glassup F4 visor [75] | 5Mp | √ | - | - | √ | GPS |
| 9 | Kopin Solos Smart Glasses [71] | - | √ | - | - | √ | GPS |

Note: ‘GAM’ represents gyroscope, accelerometer and magnetometer.

**Table 3. Computational capabilities of recent AR HMDs.**

| # | Device | Processing unit | RAM | Storage | WiFi | Blt | USB | Battery |
|---|---|---|---|---|---|---|---|---|
| 5 | Magic Leap 1 [67] | NVIDIA Parker Denver 2.0 | 8GB | 128GB | √ | 4.2 | Type-C | 8.4Wh |
| 15 | ODG R-9 [76] | Qualcomm Snapdragon 835 | 6GB | 128GB | √ | 5.0 | Type-C | 1300mAh |
| 1 | Vuzix M400 Version 1.1.4 [63] | Qualcomm Snapdragon XR1 | 6GB | 64GB | √ | 5.0 | Type-C | 135mAh |
| 3 | Microsoft HoloLens 2 [65] | Qualcomm Snapdragon 850 | 4GB | 64GB | √ | 5.0 | Type-C | 16500mAh |
| 4 | ThirdEye Gen X2 Mixed Reality Smart Glasses [66] | Qualcomm Snapdragon XR1 | 4GB | 64GB | √ | √ | Type-C | 1750mAh |
| 2 | Google Glass Enterprise Edition 2 [64] | Qualcomm Snapdragon XR1 | 3GB | 32GB | √ | 5.0 | Micro | 820mAh |
| 11 | Epson Moverio BT-350 Smart Glasses [72] | Intel Atom x5, 1.44GHz, Quad Core | 2GB | 32GB | √ | 4.1 | Micro/2.0 | 2950mAh |
| 10 | Everysight Raptor [72] | Qualcomm Snapdragon 410E | 2GB | 32GB | √ | 4.1 | Micro | 8h |
| 7 | RealWear HMT-1Z1 [69] | Qualcomm Snapdragon 625 | 2GB | 16GB | √ | 4.1 | Micro | 3400mAh |
| 13 | RealWear HMT-1 [74] | Qualcomm Snapdragon 625 | 2GB | 16GB | √ | 4.1 | Micro/Type-C | 3250mAh |
| 12 | Epson Moverio Pro BT-2200 Smart Headset [73] | TI OMAP4460 | 1GB | 8GB | √ | 4.0 | Micro/2.0 | 1240mAh |
| 16 | DAQRI Smart Glasses [77] | Intel Core m7 6th Gen (Up to 3.10 GHz) | - | 64GB SSD | √ | 4.0 | Type-C | 5800mAh |
| 14 | Glassup F4 visor [75] | ARM Cortex A9 | - | 16GB | √ | √ | Micro | 4000mAh |
| 8 | Toshiba dynaEdge AR100 Head Mounted Display[70] | Host device's CPU & internal storage | | | √ | 4.2 | 3.0 | 1050mAh |
| 6 | Kopin Golden-i Infinity Smart Screen [68] | Host device's CPU & internal storage | | | √ | 5 | Type-C | 3200mAh |
| 9 | Kopin Solos Smart Glasses [71] | Host device's CPU & internal storage | | | √ | 4.0 | Micro | 6h |

Note: 'Blt' represents Bluetooth.

**Table 4. Display capabilities of recent AR HMDs.**

| # | Device | FoV | Optics | Resolution | Refresh Rate |
|---|---|---|---|---|---|
| 2 | Google Glass Enterprise Edition 2 [64] | 80° | Binocular | 640x360 | 120Hz |
| 10 | Everysight Raptor [72] | 75° | Binocular | 1920x1080 | 30Hz |
| 3 | Microsoft HoloLens 2 [65] | 52° | Binocular | 1268x720 | 120Hz |
| 5 | Magic Leap 1 [67] | 50° | Binocular | 1920x1080 | 60Hz |
| 15 | ODG R-9 [76] | 50° | Binocular | 1920x1080 | 60Hz |
| 8 | Toshiba dynaEdge AR100 Head Mounted Display [70] | 47° | Binocular | 1280x720 | 60Hz |
| 4 | ThirdEye Gen X2 Mixed Reality Smart Glasses [66] | 42° | Binocular | 1280x720 | 60Hz |
| 16 | DAQRI Smart Glasses [77] | 30° | Binocular | 1360x768 | 90Hz |
| 11 | Epson Moverio BT-350 Smart Glasses [72] | 23° | Binocular | 1280x720 | 30Hz |
| 12 | Epson Moverio Pro BT-2200 Smart Headset [73] | 23° | Binocular (Helmet) | 960x540 | 60Hz |
| 14 | Glassup F4 visor [75] | 22° | Monocular | 640x480 | 15Hz |
| 6 | Kopin Golden-i Infinity Smart Screen [68] | 21° | Monocular | 854x480 | 60Hz |
| 7 | RealWear HMT-1Z1 [69] | 20° | Monocular (Helmet) | 854x480 | 30Hz |
| 13 | RealWear HMT-1 [74] | 20° | Monocular | 854x480 | 30Hz |
| 1 | Vuzix M400 Version 1.1.4 [63] | 17° | Monocular/Binocular | 640x360 | 60Hz |
| 9 | Kopin Solos Smart Glasses [71] | 10° | Binocular | 400x240 | 120Hz |

## 5. Conclusions

The authors classify sixteen AR HMDs available in the market since 2017 in terms of capabilities for civil infrastructure implementation. This paper compares AR HMDs performance in the areas of weight, price, product durability, sensors, connectivity, computational capacities and display capacities. Comparatively, each AR HMD has its own advantages. The weight of most AR HMDs is under the maximum limit of safety hard hats generally used in civil infrastructure projects. Civil engineers are interested in parameters such as camera resolution, processing unit and FoV, which are related to quality of information, processing speed, and working safety scenarios, respectively. According to the classification and ranking of this study, Google Glass Enterprise Edition 2, Microsoft HoloLens 2 and Everysight Raptor have the highest combined capabilities to become popular AR HMDs for civil infrastructure. Google Glass Enterprise Edition 2 has the largest FoV, which is essential in constructions, and Everysight Raptor has the second largest FoV and second largest camera resolution. Additionally, Microsoft HoloLens 2 has all-sided sensors (including depth sensor), high-resolution display, third largest FoV and relatively balanced calculation and storage performances. Microsoft HoloLens 2 is capable of 3D monitoring and has a user-friendly cross-platform toolkit for AR application (MRTK) that engineers can implement directly.

According to the current state of the art of AR HMDs in civil infrastructure, three major developments are necessary to widen their implementation in real applications: 1) technical advancements that address theoretical problems for AR users, e.g., occlusion and synchronization problems; 2) reduction of weight and price of AR HMDs along with improved performance; and 3) a unified AR platform and a benchmark model designed specifically for civil infrastructure, so that AR applications and devices available today can be compared and combined.

## Acknowledgements

The financial support of this research is provided in part by the Air Force Research Laboratory (AFRL, Grant number FA9453-18-2-0022), and the New Mexico Consortium (NMC, Grant number 2RNA6.) The conclusions of this research represent solely those of the authors.